\pdfoutput=1

\documentclass[11pt]{article}

\usepackage[preprint]{acl}

\usepackage{times}
\usepackage{latexsym}

\usepackage[T1]{fontenc}

\usepackage[utf8]{inputenc}

\usepackage{microtype}

\usepackage{inconsolata}

\usepackage{graphicx}
\usepackage{xcolor}
\usepackage{soul}
\usepackage{multirow}
\usepackage{subfigure} 
\usepackage{booktabs}
\usepackage{enumitem}
\usepackage{amsmath}

\title{That is Unacceptable: the Moral Foundations of Canceling}

\author{
 \textbf{Soda Marem Lo\textsuperscript{1,2}},
 \textbf{Oscar Araque\textsuperscript{3}},
 \textbf{Rajesh Sharma\textsuperscript{4,5}},
  \textbf{Marco Antonio Stranisci\textsuperscript{1,2}},
\\
\\
\\
 \textsuperscript{1}Università degli Studi di Torino,
 \textsuperscript{2}aequa-tech,
  \textsuperscript{3}Universidad Politécnica de Madrid,
   \\
  \textsuperscript{4}University of Tartu, Estonia,
  \textsuperscript{5}Plaksha University, India
  \\
  \small{
   \textbf{Correspondence:} \href{mailto:marcoantonio.stranisci@unito.it}{marcoantonio.stranisci@unito.it} \href{mailto:sodamarem.lo@unito.it}{sodamarem.lo@unito.it}
  }
}

\usepackage{cleveref}
\usepackage{url}

\begin{document}
\maketitle
\begin{abstract}
Canceling is a morally-driven phenomenon that hinders the development of safe social media platforms and contributes to ideological polarization. To address this issue we present the \textbf{C}anceling \textbf{A}ttitudes \textbf{De}ection (CADE) dataset, an annotated corpus of canceling incidents aimed at exploring the factors of disagreements in evaluating people canceling attitudes on social media. Specifically, we study the impact of annotators' morality in their perception of canceling, showing that morality is an independent axis for the explanation of disagreement on this phenomenon. Annotator's judgments heavily depend on the type of controversial events and involved celebrities. This shows the need to develop more event-centric datasets to better understand how harms are perpetrated in social media and to develop more aware technologies for their detection. 

\textbf{WARNING:} the CADE corpus could contain racist, sexist, violent, and generally offensive content.
\end{abstract}

\section{Introduction}\label{introduction}
The recent interest of Natural Language Processing (NLP) scholars in morality is driven by the conviction that the moral stance of people shapes their view of the world \cite{forbes2020social} and motivates their behavior \cite{van-der-meer-etal-2023-differences}. In the context of social media interaction, the pluralistic nature of morality \cite{graham2008moral} is a proxy to understand their attitudes towards the increasing amount of polarizing events~\cite{falkenberg2024patterns} that generate an escalation of violence.

The so-called Cancel Culture~\cite{Clark2020} is a representative example of how such a moral polarization works: celebrities' behaviors that are perceived as morally wrong by communities of users trigger violent reactions aimed at excluding them from the public sphere. Being able to automatically identify and mitigate this form of public shaming would be a crucial step in preserving the well-being of people in social media \cite{davani2024disentangling}. 

The main objective of our research is to provide the first study of canceling attitudes through the lens of people's different moral perspectives. To this aim, we developed the Canceling Attitudes DEtection (CADE) dataset: a corpus of canceling incidents gathered from YouTube. The corpus includes six videos regarding controversial events about celebrities and comments that have been annotated for the presence of canceling attitudes against them. Given the social relevance of the task, we involved three types of stakeholders (activists, researchers, and students) in a participatory annotation lab~\cite{Delgado2023}. Throughout the lab, we constantly received feedback from them about the annotation process and the potential uses of CADE in downstream applications for content moderation. The moral perspectives of annotators are identified according to the Moral Foundations Theory (MFT) \cite{graham2008moral}, which enables the identification of people's moral profiles by ranking their attitudes towards five foundations: care of the most vulnerable members of the community, fairness in the cooperation with others, loyalty to the group, respect of the authority, and sense of purity. By developing CADE we investigate two research questions.

\textbf{RQ1: What is the impact of an individuals' morality in evaluating canceling attitudes?}
CADE enables a systematic study of human disagreement that is not limited to individuals' sociodemographics \cite{sap2019risk,frenda-etal-2023-epic} but also considers their moral stance towards canceling attitudes. We clustered annotators on the basis of their moral profiles and observed how they evaluate YouTube comments on controversial events involving celebrities. The analysis shows that individuals' morality is event-focused: people with different moral profiles tend to evaluate differently the canceling attitudes against specific celebrities, independently of their demographic traits.

\textbf{RQ 2: Do different LLMs align with different moral profiles in evaluating canceling attitudes?} We measured the moral stance towards canceling of $6$ LLMs, which has been compared with the morality of human annotators. We compared each LLM with annotations provided by people grouped along different axes: moral profile, type of stakeholder, and gender. Our results show that different LLMs exhibit different moral perspectives when they evaluate canceling attitudes. 

This paper is organized as follows. In \Cref{sec:related} we present relevant related work; in \Cref{sec:corpus} the creation of the corpus and in \Cref{sec:corpus-analysis} its analysis. In \Cref{sec:experiments} we present two experiments relative to our two RQs. \Cref{sec:discussion} summarizes our findings. 

\section{Related Work} \label{sec:related}
\textit{Canceling} is  ``an expression of agency, a choice to withdraw one’s attention from someone or something whose values, (in)action, or speech are so offensive, one no longer wishes to grace them with their presence, time and money'' \cite{Clark2020}. The phenomenon, which originated in Black Twitter \cite{Clark2015BlackTB, Ng2022}, has evolved into a mainstream practice within internet activism that shifted towards online censorship, silencing, and aggression with growing concerns about safety in social media platforms. Currently, there is a lack of NLP works on cancel culture with the exception of a dataset generated automatically by combining sentiment analysis and Named Entity Recognition \cite{erker-etal-2022-cancel}. Our work is the first attempt to develop a manually annotated corpus on this complex phenomenon.

The growing interest of the NLP community in moral pluralism \cite{graham2008moral,schwartz2012overview} resulted in the creation of resources \cite{araque2020moralstrength,hoover2020moral} aimed at analyzing the impact of morality in perceiving harmful contents \cite{stranisci2021expression,davani2024disentangling}. Moreover, there is an active line of research that studies the morality of LLMs \cite{abdulhai-etal-2024-moral,rottger-etal-2024-political}. Our work applies moral pluralism to understand canceling attitudes in social media platforms, where the morality of LLMs can have an impact in content moderation.
\section{The CADE Dataset}\label{sec:corpus}
In this section we present the process that led to the creation of the Canceling Attitudes DEtection (CADE) dataset: a corpus of YouTube comments annotated for the study of canceling attitudes against controversial events. The section is organized as follows: we first present the data collection phase (\Cref{collection}), then we describe the design of the annotation task (\Cref{annotation-task}), and present the Annotation Lab: the participatory approach that we adopted to recruit annotators (\Cref{annotation-lab}).

\subsection{Data collection}\label{collection}
As described in \Cref{sec:related}, cancel culture is characterized by the online public shaming of celebrities for their actions or behaviors. In a NLP perspective, this can be conceived as an event-centric task \cite{chen2021event}, where the characteristics of the event orient the interest of the research. Relying on this assumption we chose YouTube 
 as source for our data collection. We leveraged the public APIs\footnote{Besides the comment, we requested the username, immediately anonymised, the date and time when the comment was posted, and the number of likes. However, the final corpus does not contain any of the collected metadata, which will not be released.}
 to obtain communicative situations where the presence of canceling attitudes can be assessed.

Firstly, we selected controversial events related to six celebrities that refer to various topics: J.K. Rowling (homo-transphobia), Kanye West (antisemitism), Lizzo (harassment), Halle Bailey (anti-woke culture), Ellen DeGeneres (bullying), Andrew Tate (sexual assault).
For each of them, we manually selected a video of a news broadcast reporting the event the target celebrity got canceled for. We limited potential biases linked to the political orientation of the source by balancing between right- and left-leaning channels, as reported in \Cref{tab:video_info} (Appendix \ref{app-annotation_materials}).
Since the annotators had to watch the video before completing the annotation task, we opted for the most viewed clip among those that did not exceed 4 minutes. For each video we extracted all the comments and randomly sampled 350 comments for each celebrity, resulting in a total of $2,100$ texts for the annotation. 

\subsection{Design of the annotation task}\label{annotation-task} 
The design of the annotation task takes into consideration two research needs: the identification of annotators' moral profiles and the definition of an annotation scheme that is effective in representing the complexity of social interactions canceling attitudes rely on.

\paragraph{Moral Foundation Questionnaire.} 
In social psychology, a common method for eliciting people's moral profile is the administration of questionnaires \cite{graham2013moral,hinz2005investigating}, which have been recently used in NLP to assess the moral stance of LLMs \cite{abdulhai-etal-2024-moral} and people \cite{davani2024disentangling}. Coherently with this research, we chose the 30-item Moral Foundation Questionnaire (MFQ30) \cite{graham2013moral}\footnote{\url{https://moralfoundations.org/questionnaires/}}, consisting of two blocks. In the first block, respondents have to reply to the question \textit{When you decide whether something is right or wrong, to what extent are the following considerations relevant to your thinking?}, rating 15 sentences on a scale from 0 (This consideration has nothing to do with my judgments of right and wrong) to 5 (This is one of the most important factors when I judge right and wrong). The second assignment is to read the 15 sentences and indicate their agreement or disagreement using a scale from 0 (strongly agree) to 5 (strongly disagree). 
By aggregating respondents' replies, it is possible to elicit which moral foundations contribute the most to their moral profile.

\paragraph{Annotation scheme.}
The annotation scheme is composed of two axes: stance \cite{aldayel2021stance} and acceptability \cite{forbes2020social}. The former is adopted to annotate the stance of the comment towards the celebrity; acceptability is adopted to evaluate whether the comment is perceived as morally unacceptable by the annotator. 

For each of the six target celebrities, we prepared a summary of the controversial event that introduced the topic to the annotators (Appendix \ref{app-event_description}) and then asked them to watch a YouTube video 
on the controversial event (Appendix \ref{app-annotation_materials}). 
Following this step, they could annotate
the stance and the unacceptability of comments about celebrities. First, annotators had to evaluate whether the YouTube user intended to attack, defend or was neutral towards the controversial event involving the target celebrity (stance).
Then annotators had to evaluate the social unacceptability of the comment, choosing on a scale that ranges from 1 (totally acceptable) to 4 (totally unacceptable). 

\subsection{Annotation Lab}\label{annotation-lab}
Since our research aims to identify how people with specific moral profiles perceive canceling attitudes in a realistic scenario, our annotator recruitment strategies focused on the involvement of people who have specific interests in the issue. We engaged with three stakeholders: activists against online discrimination, AI researchers, and NLP students. For each stakeholder we organized an annotation lab, which is designed following the literature on Participatory AI \cite{Delgado2023}: people are not only recruited as annotators, but are involved during the whole dataset creation process. For this reason, we involved stakeholders rather than relying on crowdsourcing annotator platforms. 

The annotation lab is structured in three main steps: \textit{i)} engagement: we presented a set of slides through a shared video in which we provided a definition of the phenomenon of cancel culture, we shared our research objective and its social impact and explain the whole annotation process; \textit{ii)} annotation: people were asked to fill out the MFT questionnaire and to annotate YouTube comments according to our annotation scheme; \textit{iii)} feedback: we asked participants to provide feedback about the annotation process and share their ideas about potential applications of a technology for the identification of canceling attitudes. 
Differently from traditional data annotation methodology in the NLP field, we preferred giving priority to the direct participation of the annotators by stating the purpose of the study. We wanted to make them aware of the goals of the research they contributed to, giving space to the possibility of discussing and explicitly sharing these objectives.
To this aim, we adopted two methodologies: the administration of a checkout questionnaire, and the organization of two focus groups. 

\paragraph{Feedback questionnaire.} We administered the questionnaire to all the involved stakeholders. It included 7 questions (Appendix \ref{app-checkout_qst}) about three aspects of the annotation lab: evaluating the experience in terms of its emotional impact and difficulty; providing improvements to the annotation scheme; and suggesting downstream applications of a technology trained on such a resource.

\paragraph{Focus group.} We organized one focus group aimed at students and one aimed at activists. During these meetings, we presented the preliminary results of the annotation task, and kicked off a semi-structured discussion to collect their feedback along three topics of interest: \textit{i)} evaluation of the data creation process; \textit{ii)} the relevance of morality for the task; \textit{iii)} the co-design of NLP technologies based on the dataset. 

\section{Corpus Analysis}
\label{sec:corpus-analysis}
In this section we analyse the CADE corpus.\footnote{\url{https://github.com/aequa-tech/canceling-attitudes}} We first describe the composition of annotators (\Cref{ss:process}), then the agreement between them (\Cref{ss:iaa}), finally we describe whether annotators moral profiles correlate with their sociodemographic traits (\Cref{ss:moral_profiles}).

\subsection{Annotation Process} \label{ss:process}
The annotation task involved  $57$ annotators belonging to three stakeholders: $30$ students, $17$ activists, and $10$ researchers. In addition, each annotator voluntarily shared their information about gender identity, and age, which was grouped into generations (Boomer, GenX, GenY and GenZ), nationality, ethnicity, education level and employment status.
All the sociodemographic data are reported in \Cref{tab:ann_demographics} in Appendix \ref{app-demographics}.
Having adopted a participatory approach to annotators recruitment (\Cref{annotation-lab}) that focuses on engaging with people who are interested in the phenomenon rather than hiring crowdworkers, the pool of annotators is not balanced along all sociodemographic axes. 

Each annotator is asked to annotate a subset of $210$ comments gathered from YouTube (\Cref{collection}): $35$ for each celebrity and the controversial event related to them. After cleaning unrelated comments, the final corpus includes $2,094$ texts and $11,935$ annotations. Each text has been annotated an average of $5.7$ times (with a median of $6$). Annotators had the option to refrain from assigning a label. We eliminated those who did not finish more than one-third of the task. Appendix \ref{app:annotation-statistics} reports more detailed statistics on the annotations, and Appendix \ref{app:examples} textual examples.

\subsection{Inter Annotator Agreement} \label{ss:iaa}
To assess the Inter Annotator Agreement (IAA) between annotators we employed Krippendorff's Alpha \cite{krippendorff2011computing}, which handles both agreement by chance (as the more common Cohen's Kappa agreement score), and incomplete annotations. We computed the IAA per sample and averaged them. Adopting \citeauthor{8d20e0b8-89d8-3d65-bcf5-8c19d56ec4ab}'s terminology \citeyearpar{8d20e0b8-89d8-3d65-bcf5-8c19d56ec4ab}, it resulted in a moderate agreement on stance ($0.501$), and a fair agreement on acceptability ($0.222$). These results are comparable with other datasets released in a disaggregated fashion on analogous tasks, such as offensive speech detection \cite{sachdeva-etal-2022-measuring}.
The strong difference between stance and acceptability highlights how people tend to agree more when judging YouTube users' intentions rather than when they are asked to evaluate which messages they consider unacceptable \cite{davani2024disentangling}.
Labeling 
the stance expressed by a comment results less subjective than stating 
whether the comment is 
to consider unacceptable. We report the scores broken down by sample in \Cref{tab:sample-iaa} in Appendix \ref{app-annotation_materials}.

\subsection{Annotators Moral Profiles} \label{ss:moral_profiles}
As a third part of the analysis of our annotation corpus, we assign annotators' moral profiles by leveraging the results of the MFT Questionnaire (\Cref{annotation-task}) and testing to which extent they correlate with the following sociodemographic characteristics: gender, age, and stakeholder type.\footnote{Since the recruitment of stakeholder has been performed in European countries, we decided not to include 
nationality and ethnicity as variables of our study, as they result to be unbalanced.
 We also excluded the education level and employment status because they are strongly dependent on the type of stakeholders involved during the annotation, which is the only social condition monitored by design.} 

We first computed the moral profile of annotators by obtaining their scores over the five foundations (care, fairness, loyalty, authority and purity). Each foundation is assigned a score between 0 (irrelevant) and 5 (very relevant). Together, the five scores represent the moral profile of the annotator.   

We then computed the Pearson coefficient between these profiles and gender, age, and stakeholder type. We found no statistically significant relationship between traits and moral foundations. This suggests that when annotators are grouped based on demographics, their moral profiles exhibit high variability and sparsity.

Given these preliminary results, we assigned a moral profile to each annotator through a clustering process that relies on their replies to the questionnaire. Annotators were represented with a vector with 5 dimensions, where each value in the vector represents the score obtained by the annotator for each moral foundation. 
We computed the pairwise distance with Euclidean metric and performed Agglomerative Clustering, which is preferred because it does not require setting the number of clusters as a parameter. We used Ward's linkage criterion; to choose the best number of clusters we relied on three intrinsic evaluation metrics that do not need ground truth labels, namely Silhouette Coefficient \cite{ROUSSEEUW198753}, Calinski Harabaz Index \cite{Caliński01011974} and Davies Bouldin Index \cite{Bouldin1979}. We obtained $2$ clusters, cluster\_0 (CL\_0) with $41$ annotators, and cluster\_1 (CL\_1) with $16$.

\begin{figure}
    \centering
    \includegraphics[width=0.75\columnwidth]{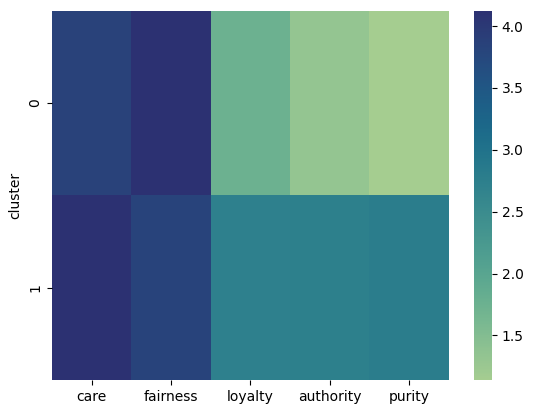}
    \caption{Heatmap of the mean scores for the five foundations across clusters.}
    \label{fig:cluster-foundations}
\end{figure}

We performed a qualitative analysis of the obtained moral clusters to understand which patterns emerge from their composition.
First, we analyzed whether the clusters were consistent with the scores that individuals obtained on each foundation. The heatmap shown in \Cref{fig:cluster-foundations} represents 
the mean score of the five foundations for each cluster. Looking at the higher divergence, we computed the Pearson coefficient between the clusters and each of the moral foundations, reporting a correlation with loyalty ($r = 0.598$ with $p-value = 9.191 \cdot 10^{-7}$), authority ($r = 0.759$ with $p-value = 7.562 \cdot 10^{-12}$) and purity ($r = 0.792$ with $p-value = 2.063 \cdot 10^{-13}$).

These results are theoretically motivated by the MFT theory, which distinguishes between individual binding foundations (care, harm), rooted on the preservation of individual freedom, and group-binding foundations (loyalty, authority, and purity) which focus on the duties of individuals towards their social groups~\cite{haidt2007morality}. In this sense, cluster\_1 is characterized by a higher moral attitude towards belonging to a group.

\begin{table}[]
    \centering\small
    \resizebox{1\columnwidth}{!}{
    \begin{tabular}{llcc}
    \hline
          \multicolumn{2}{c}{Demographics}&   CL\_0& CL\_1\\
    \hline
  \multirow{4}{*}{Gender identity} &Female& 23 (67.7\%) & 11 (32.3\%) \\
 & Male& 16 (80\%) & 4 (20\%)\\
 & Non-binary& 2 (100\%)&-\\
 & Prefer not to say& -&1 (100\%)\\
 \hline
 \multirow{4}{*}{Generation}& Boomer& 1 (100\%)&-\\
 & GenX& 1 (50\%) &1 (50\%)\\
 & GenY& 13 (73\%) & 5 (27\%)\\
          &GenZ& 
     26 (73\%)& 10 (27\%)\\
 \hline
 \multirow{3}{*}{Stakeholder}& Activist& 15 (89\%) &2 (11\%)\\
 & Researcher& 7 (70\%)&3 (30\%)\\
 & Student& 19 (64\%) &11 (36\%)\\
 \hline
 \end{tabular}}
    \caption{Composition of the clusters (CL\_0 and CL\_1) in respect to annotators' gender identity, generation and role as stakeholder.}
    \label{tab:cl-composition}
\end{table}

Once moral clusters have been validated against the MFT, we investigated again the presence of sociodemographic patterns in clusters to gain more insights from their potential correlation with moralities.
\Cref{tab:cl-composition} shows the composition of moral clusters broken down by annotators' gender, generation, and role as stakeholders. As it can be observed, the distribution of moral clusters along the gender axis is uniform: 32\% of women and 20\% of men belong to cluster\_1, showing a distribution that can be explained by annotators' gender imbalance. This is even more emphasized if generations are observed: moral clusters have the same percentage of Generation Y and Generation Z annotators, despite their age being highly unbalanced towards the latter. This suggests that, in the context of this research, age is not a factor in determining the morality of annotators. Observing the distribution of moral clusters among stakeholders shows interesting patterns, instead. If on one side 36\% of students and 30\%  of researchers belong to cluster\_1, only 11\% of activists belong to this cluster. 
\section{Experiments} \label{sec:experiments}

In this section we leverage the CADE dataset to answer the two Research Questions outlined in \Cref{introduction}. In \Cref{RQ1} we analyze human disagreement in the evaluation of canceling attitudes with a specific focus on the impact of morality in this task. In \Cref{RQ2} we study how LLMs classify canceling attitudes, analyzing whether they follow specific patterns that align them to moral or sociodemographic characteristics of annotators.
\begin{table*}[h]
    \centering
    \resizebox{1\textwidth}{!}{
    \begin{tabular}{c c c |c c |c c |c c |c c |c c | cc}  
        & \multicolumn{2}{c|}{Rowling} & \multicolumn{2}{c|}{West} & \multicolumn{2}{c|}{Lizzo} & \multicolumn{2}{c|}{Bailey}  
        & \multicolumn{2}{c|}{DeGeneres} & \multicolumn{2}{c}{Tate} & \multicolumn{2}{c}{All Celeb.} \\
        & a-d-n & un. & a-d-n & un.& a-d-n & un. & a-d-n & un.& a-d-n & un. & a-d-n & un.& a-d-n & un.\\ 
        
        moral\_0 & \textbf{.13-.71-.16} &2.12& \textbf{.20-.55-.24}&1.76& .63-.08-.26&2.72& \textbf{.40-.34-.26}&\textbf{1.77}& .67-.13-.20&\textbf{2.05}& \textbf{.14-.71-.15}&\textbf{1.78}& .37-.42-.21&2.03\\
        moral\_1 & \textbf{.21-.59-.19}&2.06& \textbf{.27-.47-.25}&1.71& .63-.06-.30&2.59& \textbf{.35-.32-.32}&\textbf{1.90}& .66-.13-.19&\textbf{2.27}& \textbf{.15-.61-.23}&\textbf{1.66}& .38-.37-.25&2.03\\
        \midrule
        
        men & .17-.67-.15&\textbf{2.05}& .24-.52-.23&\textbf{1.67}& .63-.09-.25&2.64& .40-.34-.25&\textbf{1.66}& \textbf{.66-.15-.17}&\textbf{2.03}& \textbf{.13-.74-.13}&1.77& .38-.42-.20&1.96\\
        women & .14-.68-.17&2.15& .21-.53-.25&1.79& .63-.07-.29&2.72& .38-.32-.30&\textbf{1.90}& \textbf{.67-.11-.21}&\textbf{2.17}& \textbf{.15-.66-.19}&1.73& .36-.40-.24&2.07\\
        \midrule
        
        activist & \textbf{.12-.74-.13}&2.15& .22-.53-.23&\textbf{1.86}& \textbf{.63-.12-.22}&2.77& .38-.33-.27&\textbf{1.89}& .65-.13-.20&2.07& .13-.72-.14&\textbf{1.82}& .36-.44-.20&2.09\\
        student & \textbf{.18-.63-.19}&2.11& .21-.54-.25&\textbf{1.72}& \textbf{.63-.05-.30}&2.63& .40-.33-.27&1.81& .66-.12-.20&\textbf{2.16}& \textbf{.15-.64-.20}&\textbf{1.73}& .37-.38-.25&2.02\\
        researcher & .13-.72-.15&\textbf{2.00}& .26-.49-.24&\textbf{1.61}& .63-.09-.28&2.66& .34-.34-.32&\textbf{1.67}& .68-.14-.17&\textbf{2.04}& .13-.73-.14&1.67& .36-.42-.21&2.04\\
        \midrule
        
        all & .15-.68-.17&2.1& .23-.53-.24&1.7& .64-.08-.28&2.6& .39-.33-.28&1.8& .67-.13-.20&2.1& .14-.69-.17&1.7& &\\
        \bottomrule
    \end{tabular}}
    \caption{Percentage of Attack (a), Defend (d) and Neutral (n) labels, and the average unacceptability (un.) for each group and celebrity. In bold scores with a chi-square test below $p<0.05$. Only for stakeholders, the chi-squared is computed one vs. all.}
    \label{tab:stance-acc_all}
\end{table*}

\subsection{STUDY 1: What is the impact of individuals' morality in evaluating canceling attitudes?}\label{RQ1}
The first experiment aims to analyze the human disagreement of annotators in perceiving canceling attitudes, focusing in particular on the influence of their moral profiles (\Cref{ss:moral_profiles}) in their evaluation. Given the type of dataset, which replicates online situations triggered by controversial events, we analyze to which extent human disagreement is determined by the type of events and celebrities targeted by YouTube comments.

The first step of the study is the analysis of stance and unacceptability emerging from YouTube comments, broken down by celebrity. \Cref{tab:stance-acc_all} shows the annotations aggregated by celebrities (the columns) and moral and sociodemographic groups (the rows). For each celebrity, the relative distribution of stance labels (a-d-n) and the average unacceptability scores are reported.

The table shows that \textbf{canceling attitudes vary significantly among celebrities}. Looking at the last row, only a few comments attack J.K. Rowling ($15\%$) for her transphobic declarations and Andrew Tate ($14\%$) for the accusation against him of sexual assault. Conversely, Lizzo suffers a high number of attacks ($64\%$) for her misconduct against a member of her staff, as well as Ellen DeGeneres ($67\%$) for her bullying attitudes in her working environment. 
Observing the correlation between the level of unacceptability and stance (\Cref{tab:distr-stance-unaccept}), it emerges the tendency for attacking to correspond with increased unacceptability, especially compared to a neutral stance. Defend presents more varied results, with a notable percentage of cases showing moderate levels of unacceptability (2 and 3). However, when looking at the celebrity level (\Cref{tab:stance-acc_all}), the perceived unacceptability of YouTube comments appears to be orthogonal to stance. Comments can be perceived as highly unacceptable regardless they attack celebrities or not. For instance, J.K. Rowling is mostly defended but the unacceptability score is high. Since annotators were asked how much a comment contributed to shaming a target, we infer that such a result suggests many comments defend her by adopting canceling attitudes against others: in these cases Daniel Radcliffe and Emma Watson, who called out against her transphobic claims. Lizzo is mostly attacked and comments about her scored the highest unacceptability ($2.6$): this means that she is the actual target of canceling attitudes. Comments on Andrew Tate and Kanye West, who mainly defend them, also score the lowest unacceptability scores ($1.7$), showing that their focus is more on defending them rather than attacking other targets. 


\begin{table}
    \centering
    \small
    \begin{tabular}{ccc}
    \hline
        Unacc. & Stance & \% Annotations \\ 
        \hline
        \multirow{3}{*}{1} & Attack & 15.97 \\ 
        ~ & Defend & 52.80 \\ 
        ~ & Neutral & 31.23 \\ 
        \hline
        \multirow{3}{*}{2} & Attack & 41.46\\ 
        ~ & Defend & 39.36 \\ 
        ~ & Neutral & 19,18 \\ 
        \hline
        \multirow{3}{*}{3} & Attack & 52.36 \\
        ~ & Defend & 35.43\\ 
        ~ & Neutral & 12.21\\
        \hline
        \multirow{3}{*}{4} & Attack & 74.61\\
        ~ & Defend & 18.35 \\ 
        ~ & Neutral & 7.04\\ 
    \hline
    \end{tabular}
    \caption{Distribution of texts labeled as Attack, Defend, and Neutral, broken down by the range of unacceptability. The final column reports the percentage of texts for each stance within each acceptability level, relative to the total number of texts at that acceptability level. As annotators could abstain from labeling, results are computed based on the instances with completed annotations for both tasks, thus excluding $350$ annotations.}
    \label{tab:distr-stance-unaccept}
\end{table}

The analysis of annotations broken down by moral cluster, gender, and stakeholder shows lower variation among moral and sociodemographic axes than one emerging from the analysis of celebrities (\Cref{tab:stance-acc_all}, last column). The percentage of comments annotated as attacking celebrities is uniform along morality, gender, and type of stakeholder, ranging from $0.36$ to $0.38$. The percentage of comments evaluated as defending celebrities shows more variation along moral clusters (cluster\_0: $42\%$ \textit{versus} cluster\_1: $37\%$) and stakeholders (activists: $44\%$ \textit{versus} students: $38\%$). Variation in the evaluation of unacceptability is even less significant. Annotators' moral profiles appear to have no impact on the perception of this phenomenon: the average unacceptability is $2.03$ for both groups. Stakeholders show some variation, with the average unacceptability provided by activists ($2.09$) that diverges by $0.07$ points from the average unacceptability provided by students ($2.02$). Differences along the gender axis are more significant, as it is possible to observe through the variation between the unacceptability score assigned by women ($2.07$) and the one assigned by men ($1.96$). 

If the behaviors of different moral and sociodemographic groups are observed through the lens of specific events, very diverging patterns in the evaluation of canceling attitudes emerge. 

For each axis of interest (e.g., morality) we computed the relative distribution of stance and the average unacceptability scores and computed the Chi-squared test between the distribution of labels assigned by annotators belonging to different groups (e.g., cluster\_0 \textit{vs} cluster\_1) in order to assess if there is a significant divergence between their evaluations. All the cases where the Chi-squared test shows a statistically significant divergence ($p<0.05$) are highlighted in bold in \Cref{tab:stance-acc_all}. As it can be observed in the table, the different perception of stance among annotators with different moral profiles clearly emerges if specific celebrities are examined. The two moral clusters diverge in the evaluation of attacking comments against J.K. Rowling (cluster\_0: $13\%$ \textit{versus} cluster\_1: $21\%$) and Kanye West (cluster\_0: $20\%$ \textit{versus} cluster\_1: $27\%$) and of defending comments against Andrew Tate (cluster\_0: $71\%$ \textit{versus} cluster\_1: $61\%$). Similar variations can be observed at the level of the gender axis and stakeholder axis. Women and men diverge the most in evaluating the defending stance on comments on Andrew Tate (men: $74\%$ \textit{versus} women $66\%$); students ($18\%$) perceive more attacking comments against J.K. Rowling than activists ($12\%$) and researchers ($12\%$).

The perception of unacceptability shows interesting patterns as well. Except for comments about Lizzo, where there is no statistically significant divergence along any considered axis, each axis of moral and sociodemographic variation shows differences in the attribution of the unacceptability of comments. In 5 cases out of 6 women and activists rank YouTube comments as more unacceptable on average. In 4 cases out of 6 people belonging to the moral cluster\_0, which is the one more tied to individual bindings, are inclined to label comments as more unacceptable.

The study shows that the magnitude of canceling attitudes significantly varies depending on the type of controversial event: some celebrities are more likely to trigger violent reactions than others. Human disagreement in the evaluation of canceling attitudes in YouTube comments heavily depends on such variation. This is particularly true for the role of morality in evaluating unacceptability. When patterns of annotation towards specific celebrities are observed, \textbf{morality appears to drive the disagreement between annotators} more than their gender and stakeholder type.


\subsection{STUDY 2: Do Different LLMs Align with Different Moral Profiles in Evaluating Canceling Attitudes?}\label{RQ2}

\begin{figure*}
    \centering
    \includegraphics[width=0.9\linewidth]{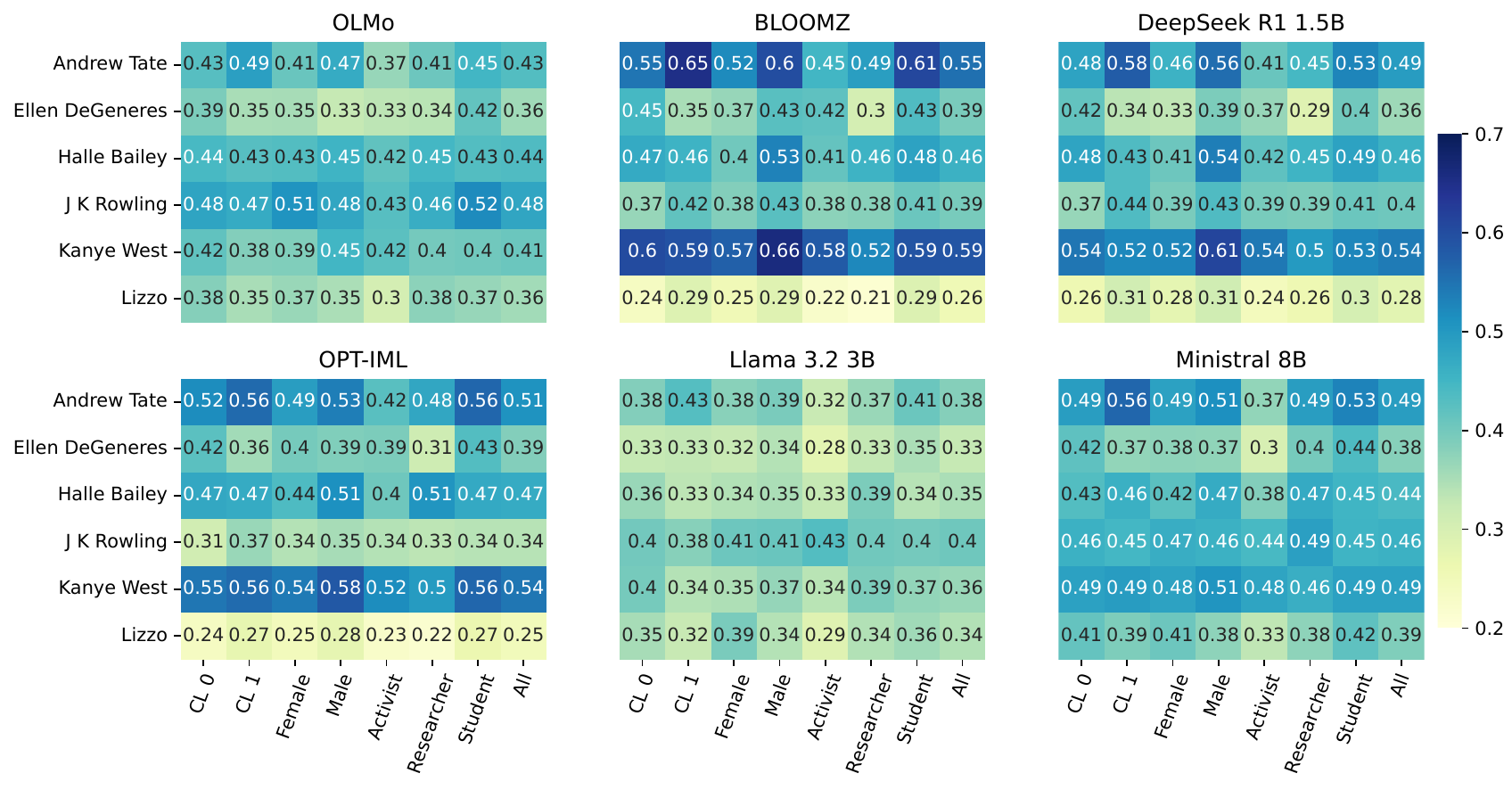}
    \caption{LLM variability in alignment on the unacceptability task. The color bar represents F1-scores, with darker shades indicating higher values}.
    \label{fig:llm-heatmap}
\end{figure*}

In this second study, we examine the attitude of LLMs towards cancel culture by observing how their performance varies in relation to human annotators.
We focus on the unacceptability task, as it is the most sensible in terms of morality, and is the target of the assessment towards cancel culture by both human annotators and language models.
To do so, we select six relevant models, similarly as done by \citet{NEURIPS2023_a2cf225b}, namely:
\texttt{OLMo 7B}, 
\texttt{BLOOMZ 3B}, 
\texttt{DeepSeek R1 1.5B}, 
\texttt{OPT-IML 1.3B}, 
\texttt{Llama 3.2 3B}, and 
\texttt{Ministral 8B}.

For extracting the models' annotations, we adopt a straightforward zero-shot setting, prompting the models to classify the unacceptability of each comment.
To analyze the generated annotations, we evaluate through a classification task, aggregating human annotations through majority vote, and treating them as a sort of gold standard.
Although the models are not actually performing a classification task, we framed the evaluation as such.
In this way, we use the macro-averaged F1-score to evaluate language models' performances using human annotations as a reference.

Figure~\ref{fig:llm-heatmap} shows the result of this experiment.
On an aggregate level, we observe that the models largely vary their alignment to human annotators when considering the different celebrities.
These variations show when examining both celebrities (vertical axis) and annotator groups (horizontal axes).
In light of this observation, we argue that language models show a similar behavior to that of human annotators, varying their perception towards cancel culture with the event that is being assessed.

From an in-depth observation of \Cref{fig:llm-heatmap} both common and specific patterns among models emerge. LLMs show different degrees of alignment with specific moral clusters. OLMo classifies unacceptability with a highest F-score against annotations provided by moral cluster\_0 in $5$ cases out of $6$ while OPT-IML shows an opposite alignment, obtaining a highest F-score against moral cluster\_1 annotations in $4$ cases out of $5$. The alignment of LLMs with annotators broken down by gender appears to systematically lean towards men. BLOOMZ and DeepSeek always obtain the highest F-1 scores against men's annotations, OPT-IML in $5$ cases out of $6$. Models that align the most with women are OLMo and Ministral, which both score a highest F-score against them in $3$ cases out of $6$. The analysis of stakeholders shows a general higher alignment of LLMs with students, which are the most represented group in our pool of annotators. In particular, BLOOMZ always aligns with this category of stakeholders. On the opposite, activists are the stakeholders against which models are more misaligned. Specifically, Llama and Ministral obtain the lowest F-score against annotations provided by this group in $5$ cases out of $6$.

Some general patterns seem to emerge if specific celebrities are observed. All models show the highest agreement with men and annotators belonging to the moral cluster\_1 in the evaluation of unacceptability of comments against Andrew Tate, and the lowest agreement with activists. Other general patterns are specific to the alignment of models with moral clusters. $5$ models out of $6$ align with moral cluster\_0 in the evaluation of unacceptability against Ellen DeGeneres and Halle Bailey, while the remaining models obtain the same F-1 score for both moral clusters. LLMs' classifications of unacceptability of comments against Halle Bailey and Kanye West always lean towards men's annotations while there are no celebrities against which models systematically align with women in the classification of unacceptability.

In terms of comparing the annotator groups, we see that there are differences between the moral clusters, being BLOOMZ the model that shows the largest variation.
Likewise, we observe differences between the female and male groups, with several models showing greater alignment with the male group.
A relevant exception is Ministral, that shows a balance behavior in terms of gender, obtaining high scores overall.
Focusing on the stakeholders group, we also observe more variation in this evaluation.
In general, the models align with the student group to a greater extent.

The study shows that \textbf{LLMs tend to exhibit different perspectives in classifying unacceptability}. By observing their F-score on each event, it is possible to notice models that are more aligned with a specific moral cluster in comparison to models that are more aligned with the other one. Models also diverge along the gender axis: some of them always lean towards men's annotation while others show variation in their alignment with men and women. As for human annotations, the type of event influences the alignment between groups of annotators and LLMs classification, which in some cases systematically leans towards specific groups while in others it triggers divergent behaviors between models.

\section{Discussion} \label{sec:discussion}

A first result that emerges from our study is that \textbf{canceling attitudes heavily depend on the types of controversial event that triggers them}. YouTube comments towards certain celebrities are systematically considered less acceptable by all annotators, regardless of their background. This demonstrates that the so-called Cancel Culture does not have the same impact on all its victims.

This event-based variation explains human disagreement in the evaluation of canceling attitudes. Rather than operating at a general level, \textbf{morality, gender, and stakeholder type drive the disagreement of annotators against specific celebrities}. In this context, morality appears to have a significant influence, especially in the evaluation of unacceptability. In $4$ cases out of $6$, annotators with different moral profiles show a statistically significant variation (\Cref{tab:stance-acc_all}) in judging unacceptability. This highlights the importance of developing resources that better represent the communicative context in which potentially harmful content is spread. 


The analysis of LLMs behavior in recognizing canceling attitudes shows similar patterns to the ones observed in the study of human disagreement. LLMs align with specific groups about certain celebrities. Additionally, 
\textbf{different models tend to align with different categories of annotators characterized by their sociodemographic characteristics and morality}. Controlling these patterns of alignment would be a significant step in the implementation of fairer technologies for reducing the ideological polarization between users on social media platforms.  

\section{Conclusion}
In this paper, we presented the Canceling Attitude DEtection (CADE) dataset, a corpus specifically designed to investigate how people with different moral views of the world evaluate this phenomenon. The corpus, which includes six canceling incidents gathered from YouTube, has been annotated by annotators belonging to three categories of relevant stakeholders for this social issue (activists, researchers, and students) whose moral profiles have been obtained through the MFT questionnaire. 

Future work will be devoted to expanding the study of canceling attitudes by including different categories of targets (e.g., organizations), as well as an extensive analysis of how unacceptability varies in respect to the stance, together with used linguistic strategies. Additionally, we will adopt an intersectional approach to better understand how morality intersects with other sociodemographic factors in determining annotators' disagreement, and to which extent LLMs are able to grasp this complex issue. To this end, we plan to analyse differences between models of the same family with a varying number of parameters, conducting extensive research on LLMs alignment.

\section*{Limitations}
While this work moves towards the development of participatory approaches to Natural Language Processing, the annotator pool is not balanced, especially leaning towards the White and Educated population. Moreover, although the chosen events had worldwide coverage, the majority of the annotators do not come from the same social background as the target celebrities. This made it possible to carry out the experiment with people from different nationalities, who, however, experienced cancel culture more as spectators than actors. In the future, we plan to expand the annotation lab to a more diverse group of annotators along all the sociodemographic axes. 
In acknowledging the low agreement resulting from the annotation in Section \Cref{ss:iaa}, we plan to compare these results with other disaggregated corpora, adopting metrics that emphasize the emergence of disagreement factors \citep{passonneau2014benefits}.
Finally, the current design involved watching only one video per celebrity, with exposure to the framing implicit in the political orientation of the chosen news source. While we aimed to ensure political diversity across the six selected videos, in the future, we intend to achieve this within the same canceling event.

\section*{Ethical Statement}
This research relies on the voluntary work of those who participated in the Annotation Labs. All the involved annotators freely accepted to take part to the laboratory, for which no compensation was provided. 
We adopted all the measures to protect data privacy and safeguard personal information of both YouTube users whose comments were collected, and annotators who participated to the task. The work has been approved by the Ethics Committee of the institution of one of the authors. 

We acknowledge that sharing the research objectives in advance can be a source of bias. However, we decided to prioritise the annotators' awareness about the technology they were building, providing an overview of the work. We made an effort to ensure the direct participation of those who contributed to the construction of the dataset, not only through annotations but also with their feedback, trying to limit the negative aspects of this choice.

In the future, we plan to expand this work to a less Eurocentric context, concerning both the chosen celebrities and the involved annotators, looking at it as a necessary improvement to foster diversity. 

\section*{Author Contribution Statement}

\begin{itemize}
    \item \textbf{Project conception:} Soda Marem Lo, Marco Antonio Stranisci.
    \item \textbf{Lead:} Soda Marem Lo.
    \item \textbf{Supervision:} Marco Antonio Stranisci.
    \item \textbf{Experiments with LLMs:} Oscar Araque.
    \item \textbf{Research design and paper writing:} Oscar Araque, Soda Marem Lo, Rajesh Sharma, Marco Antonio Stranisci.
\end{itemize}


\section*{Acknowledgements}

Oscar Araque would like to acknowledge the funding of
funded by the Spanish Ministry for Economic Affairs and Digital Transformation and by the European Union - NextGenerationEU within the Programme UNICO I+D Cloud (TSI-063100-2022-0002);
as well as the funding of the project CPP2023‐010437 financed by the MCIN / AEI /
10.13039/501100011033 / FEDER, UE.
This work has been supported by the Madrid Government (Comunidad de Madrid-Spain) under the Multiannual Agreement 2023-2026 with Universidad Politécnica de Madrid in the Line A, Emerging PhD researchers (project MORA, DOCTORES-EMERGENTES-24-9UMLXZ-37-IIGW).

The work of Soda Marem Lo was partially supported by “HARMONIA” project - M4-C2, I1.3 Partenariati Estesi - Cascade Call - FAIR - CUP C63C22000770006 - PE PE0000013 under the NextGenerationEU programme and by aequa-tech.

We thanks students of Language Technologies and Digital Humanities and Hate Trackers activists for their participation in the annotation lab.
\bibliography{latex/custom}

\clearpage

\appendix
\section{Annotator's demographics}\label{app-demographics}

\Cref{tab:ann_demographics} shows the demographic information about the annotators. We additionally asked them to indicate their nationality and first language. 38 are Italian, 2 are Spanish, 2 are Chinese, and there is one person for each of the following nationalities: Italian-Argentinian, Italian-Romanian, Iranian, Greek, Russian, Kazakhstan, Indian, Moldovan, Persian, Dutch, Romanian. As regards their mother tongue: 38 chose Italian, 3 Spanish, Russian, Chinese, 2 Persian, Romanian and Greek, and 1 Hebrew, Hindi, Farsi and Dutch.

\begin{table}[]
    \centering\small
    \begin{tabular}{lc|c}
    \hline
          \multicolumn{2}{c}{Demographics}&  \#Annotators\\
          \hline
          \multirow{4}{*}{Gender identity}&Female& 
    34\\
 & Male&20\\
 & Non-binary&2\\
 & Prefer not to say&1\\
 \hline
 \multirow{4}{*}{Generation}& Boomer&1\\
 & GenX&2\\
 & GenY&18\\
 & GenZ&36\\
 \hline
 \multirow{5}{*}{Ethnicity}& White&48\\
 & Asian&5\\
 & Mixed&2\\
 & Black&1\\
 & Caucasian&1\\
 \hline
 \multirow{4}{*}{Education level}& Bachelor degree&29\\
 & Master degree&18\\
 & Doctorate degree&5\\
 & High school diploma&5\\
  \hline
 \multirow{8}{*}{Employment}& Unemployed&16\\
 & Full-time&15\\
 & Part-time&10\\
 & Due to start a new job&3\\
 & Student&8\\
 & Not in paid work&2\\
 & Occasional work&1\\
 & Freelancer&2\\
 \hline
 \end{tabular}
    \caption{Sociodemographic information.}
    \label{tab:ann_demographics}
\end{table}

\section{Instructions for the annotation process}\label{app-event_description}
\begin{figure*}
    \centering
    \includegraphics[width=1\linewidth]{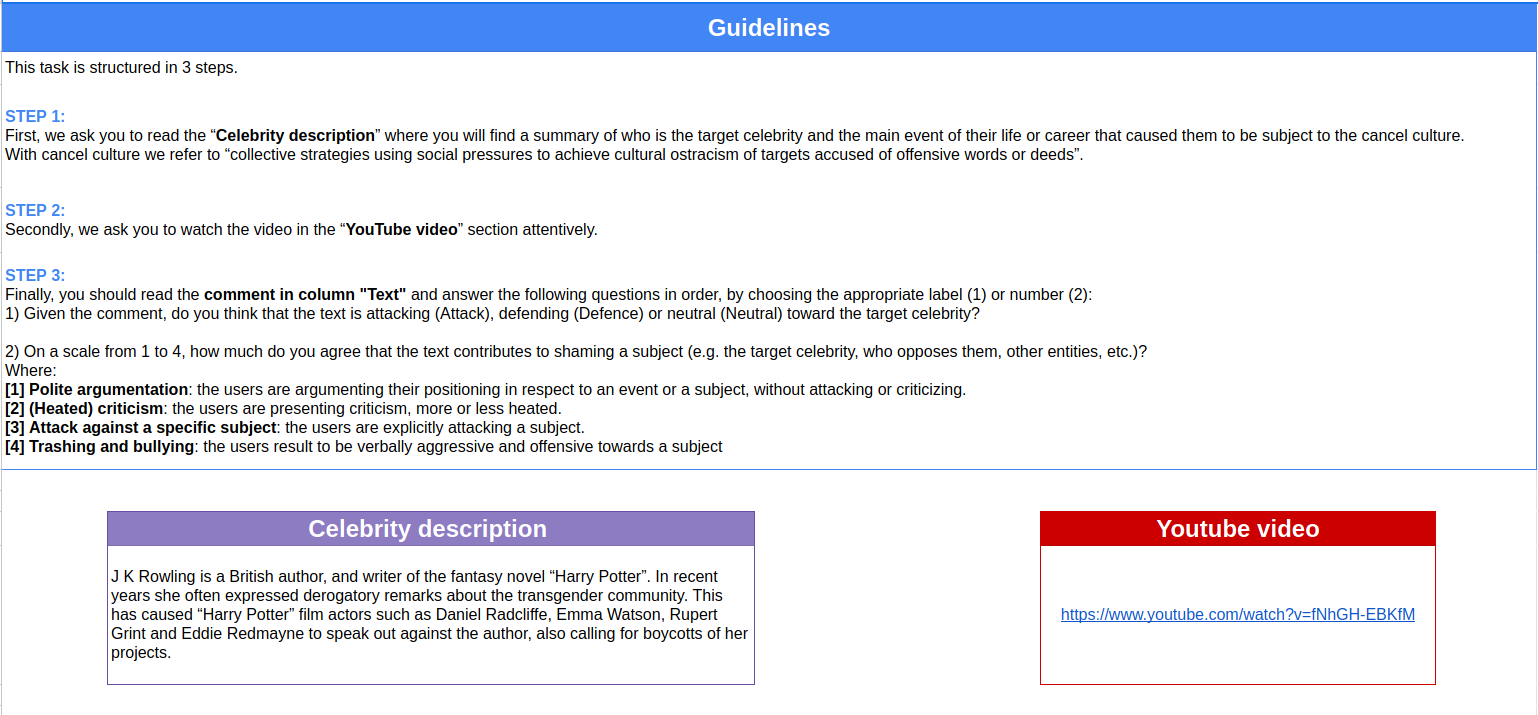}
    \caption{Instructions for the annotators.}
    \label{fig:annotation_platform}
\end{figure*}

\Cref{fig:annotation_platform} shows the instructions provided to the annotators. 

In the following, we report all the celebrity descriptions.

\paragraph{J K Rowling} J K Rowling is a British author, and writer of the fantasy novel “Harry Potter”. In recent years she often expressed derogatory remarks about the transgender community. This has caused “Harry Potter” film actors such as Daniel Radcliffe, Emma Watson, Rupert Grint and Eddie Redmayne to speak out against the author, also calling for boycotts of her projects.
\paragraph{Kanye West} Kanye West, also known as Ye, is an American rapper. He has frequently spoken out on political and social issues with controversial opinions on topics such as abortion, capital punishment, welfare and gun rights. On frequent occasions he expressed antisemitic thoughts, stating his admiration for Adolf Hitler, denying the Holocaust, and supporting other conspiracy views against Jewish people, which led him to being banned from Twitter for 8 months.
\paragraph{Lizzo} Lizzo is an American rapper and singer. Throughout her career, she has been publicly interested and outspoken on social issues. She supported the LGBTQ+ community considering herself an ally, and advocated for body positivity, being subject to body shaming herself. In August 2023, she was accused of sexual, religious and racial harassment, disability discrimination, assault, weight-shaming and a hostile work environment by three former backup dancers, supported by other co-workers.
\paragraph{Halle Bailey} Halle Bailey is an American singer and actress. In 2023 she performed as the protagonist in the Disney movie “The Little Mermaid”, a choice that was subject to widespread criticism because in the cartoon the little mermaid was depicted as white, while Bailey is black. At the time, the hashtag \#NotMyAriel was launched, leading to a discussion about Disney film revision in the name of woke culture.
\paragraph{Ellen DeGeneres} Ellen DeGeneres is an American comedian and television host, famous for “The Ellen DeGeneres Show”. In July 2020 ten former employees of this show accused her of creating a toxic environment, with racist micro-aggressions, intimidation, abuse and sexual harassment episodes against female employees. She publicly apologized, promising that she would correct the issue.
\paragraph{Andrew Tate} Andrew Tate is an American and British former professional kickboxer, who became famous for promoting misogynist and violent messages, representative of the manosphere community. He was deplatformed from Twitter, Instagram, Facebook and TikTok. His account on Twitter was reinstalled in November 2022 after the Elon Musk acquisition. In December 2022 he was arrested with charges of rape, human trafficking and forming a criminal gang for the sexual exploitation of women. He was released a few months later.
\section{Annotation materials}\label{app-annotation_materials}
\Cref{tab:video_info} presents all the information about the selected YouTube videos.

\begin{table*}[ht]
    \centering\small
    \begin{tabular}{ccccc}
\hline
Celebrity&  Topic &News Broadcast & Political orientation &Minutes\\
\hline
 J K Rowling& Homo-transphobia& Sky News Australia & Right wing & 2.29\\
 Kanye West& Antisemitism& Fox 11 Los Angeles& Right wing & 1.14\\
 Lizzo& Harassment& Fox News& Right wing &3.24\\ 
 Halle Bailey& Incelism& CBS Media& Left wing &1.47\\
 Ellen DeGeneres& Bullying& CBS Media& Left wing &3.01\\
 Andrew Tate& Sexual assault & Law and Crime network& Left wing &2.26 \\
 \hline
    \end{tabular}
    \caption{Information about the selected YouTube videos their duration for each target celebrity.}
    \label{tab:video_info}
\end{table*}

\Cref{tab:sample-iaa} reports the detailed sample composition and Inter Annotator Agreement.

\begin{table*}[]
    \centering\small
    \begin{tabular}{ccllllll}
    \hline
 Samples &\#Researchers&\#Students& \#Activists& \#Texts&\#Annotations& $\alpha$ Stance& $\alpha$ Unaceptability\\
 \hline
         Sample\_0&    1&4& 2&  210&1,470& 0.518&0.177\\
 Sample\_1&  1&2& 2&  210&1,050& 0.472&0.306\\
 Sample\_2&  1&3& 2&  208&1,248& 0.39&0.189\\
 Sample\_3&  1&3& 3&  210&1,470& 0.548&0.153\\
 Sample\_4&  1&4& 1&  209&1,254& 0.491&0.183\\
 Sample\_5&  1&4& 2&  210&1,470& 0.601&0.283\\
 Sample\_6&  1&2& 1&  210&840& 0.505&0.202\\
 Sample\_7&  1&1& 1&  210&630& 0.425&0.215\\
 Sample\_8&  1&3& 1&  208&1,040& 0.463&0.271\\
 Sample\_9&   1&4& 2&  209&1,463& 0.603&0.245\\
 \hline
 Total & 10& 30& 17&  2,094&11,935&  & \\
 $\alpha$ $\text{mean}_{\text({std})}$ & & & &  &&$0.501_{(0.069)}$&$0.222_{(0.051)}$\\
 \hline
    \end{tabular}
    \caption{Sample composition, annotation and IAA. }
    \label{tab:sample-iaa}
\end{table*}

\section{Checkout Questionnaire}\label{app-checkout_qst}
We report the questions asked to the annotators in the checkout questionnaire:

\begin{itemize}[noitemsep]
    \item What opinion do you have about the topic of \textit{social media shaming} after the annotation? 
    \item How did you find the annotation task? (difficult? emotionally impactful? etc.)
    \item Is there anything you would like to add?
    \item Is there anything you would like to change?
    \item Did the moral questionnaire influence the way you performed the annotation? 
    \item Did you perceive the completion of the questionnaire and annotation as thematically related?
    \item How would you use a tool that can recognize social media shaming? 
\end{itemize}

This questionnaire was not mandatory and received $33$ answers. Overall, the majority of annotators found that the annotation topic was related to the Moral Foundations Theory and eight of them wrote that they have been influenced by filling the MFT questionnaire before the annotation task. In a future iteration of this work we will administer the questionnaire after the annotation. A second relevant aspect emerging from the checkout survey is that some comments are not easy to be interpreted for non-native English speakers since the context in which the controversial events had place is strongly US-centric. We will integrate the existing corpus with annotations focused on incidents from different cultural contexts. Results of the checkout questionnaire are available in the project folder\footnote{\url{https://github.com/aequa-tech/canceling-attitudes/blob/main/questionnaires/checkout_questionnaire.csv}}
\section{Corpus statistics}\label{app:annotation-statistics}

The stance of the comment was labeled as Neutral $2,626$ times, Attack $4,363$, and Defend $4,821$ times. Regarding the acceptability, going on a scale from 1 (totally acceptable) to 4 (totally unacceptable), the labels were assigned as follows: [1] Polite argumentation $4,539$ times, [2] (Heated) criticism $3,532$ times, [3] Attack against a specific target $1,990$ times, [4] Trashing and bullying $1,477$ times.


\section{Example of annotated comments}\label{app:examples}

\Cref{tab:text-examples} shows a comment for each possible combination of the two annotation dimensions (stance and unacceptability).

\begin{table*}[!ht]
    \centering
    \small
    \renewcommand{\arraystretch}{1.2}
    \setlength{\tabcolsep}{5pt}
    \begin{tabular}{ccc|p{10.2cm}}
    \hline
    Stance & Unacc. & Celebrity & Text \\
    \hline
    \multirow{4}{*}{Attack} 
        & 1 & Ellen DeGeneres & So many things that go behind the scenes \\
        & 2 & Lizzo & Ain't that the pot calling the kettle black \\
        & 3 & Ellen DeGeneres & Cancel her\\
        & 4 & Ellen DeGeneres & I never liked Ellen, she is a stupid overhyped KAREN\\
    \hline
    \multirow{4}{*}{Defend} 
        & 1 & Andrew Tate & Maybe AI voice? \\
        & 2 & Andrew Tate & The Matrix is after him—“The Matrix” being others jealous of him. \\
        & 3 & Kanye West & Speaking out against the Synagogue of Satan will never end good\\
        & 4 & J.K. Rowling & 85\% of the world agrees with JK Rowling. The other 15\% have mental health issues. \\
    \hline
    \multirow{4}{*}{Neutral} 
        & 1 & Ellen DeGeneres & Bro, I wasn't even born in 2003. The show is older than me \\
        & 2 & Halle Bailey & Race swapping is racist. \\
        & 3 & Lizzo & They have you so duped!! Strangely at the same time they posted this all other stations were posting the the new Trump court orders. They steer you and control you. \\
        & 4 & Lizzo & Is my mind playing tricks on me or is the dude's head in the pin-stripe suit kind of big for his body? \\
    \hline
    \end{tabular}
    \caption{Examples of annotated texts across all combinations of stance and unacceptability. Labels reflect full or majority agreement among annotators.}
    \label{tab:text-examples}
\end{table*}

\section{LLM implementation details}

In this work, we have used the following public models:
\begin{itemize}
    \item \texttt{OLMo 7B}\footnote{\url{https://huggingface.co/allenai/OLMo-2-1124-7B-Instruct}}
    \item \texttt{BLOOMZ 3B}\footnote{\url{https://huggingface.co/bigscience/bloomz-3b}}
    \item \texttt{DeepSeek R1 1.5B}\footnote{\url{deepseek-ai/DeepSeek-R1-Distill-Qwen-1.5B}}
    \item \texttt{OPT-IML 1.3B}\footnote{\url{https://huggingface.co/facebook/opt-iml-max-1.3b}}
    \item \texttt{Llama 3.2 3B}\footnote{\url{https://huggingface.co/meta-llama/Llama-3.2-3B-Instruct}}
    \item \texttt{Ministral 8B}\footnote{\url{https://huggingface.co/mistralai/Ministral-8B-Instruct-2410}}
\end{itemize}

To generate the annotations with these models, we have followed a zero-shot approach, prompting the models to generate their annotations.
All models were executed in a NVIDIA Titan X Pascal GPU, with 12GB of memory.

To generate the annotations regarding stance, the used prompt is as follows:
\textit{Classify the text into being defensive, neutral or attacking.}
Similarly, we used the following prompt to generate the acceptability annotations:
\textit{Classify the text into a scale from 1 to 4, considering how much the text contributes to shaming or degrading a subject, being 1 the lower and 4 the higher.}

\end{document}